\documentclass[summary]{ursi}

\newcommand{\mdash}{---}
\usepackage{journals}
\newcommand{\sstng}{\textbf{SST\textsc{ng}}}
\title{Solar Submillimeter Telescope next generation}

\author{C. Guillermo Giménez de Castro*\affref{uno}\affref{dos},
  Jean-Pierre Raulin\affref{uno}, Adriana Valio\affref{uno}, Emilia
  Correia\affref{uno}\affref{tres}, Paulo J. A. Simões\affref{uno}\affref{cuatro} and Sérgio
  Szpigel\affref{uno}}

\affiliation{%
  \aff{uno}{Centro de Rádio Astronomia e Astrofísica
  Mackenzie, EE, UPM, São Paulo, Brazil; e-mail:
  guigue@craam.mackenzie.br; raulin@craam.mackenzie.br;
  avalio@craam.mackenzie.br, ecorreia@craam.mackenzie.br, paulo@craam.mackenzie.br;
  szpigel@craam.mackenzie.br}
  \aff{dos}{Instituto de Astronomía y Física del Espacio, UBA / CONICET, Buenos Aires, Argentina}
  \aff{tres}{Instituto Nacional de Pesquisas Espaciais, INPE, São José dos Campos, Brazil}
  \aff{cuatro}{SUPA School of Physics and Astronomy, University of Glasgow, Glasgow, G12 8QQ, UK}
  }
\begin{document}

\maketitle

\begin{abstract}
The Solar Submillimeter Telescope (SST) is an unique instrument that
has been observing the Sun daily since 2001 bringing a wealth of
information and raising new questions about the particle acceleration
and transport, and emission mechanisms during flares.  We are now
designing its successor, the \sstng, that will expand the scientific
goals of the instrument, including non-solar source observations.

\end{abstract}

\section{Introduction}

\begin{figure}[htbp]
  \centering
  \centerline{\includegraphics[width=40mm]{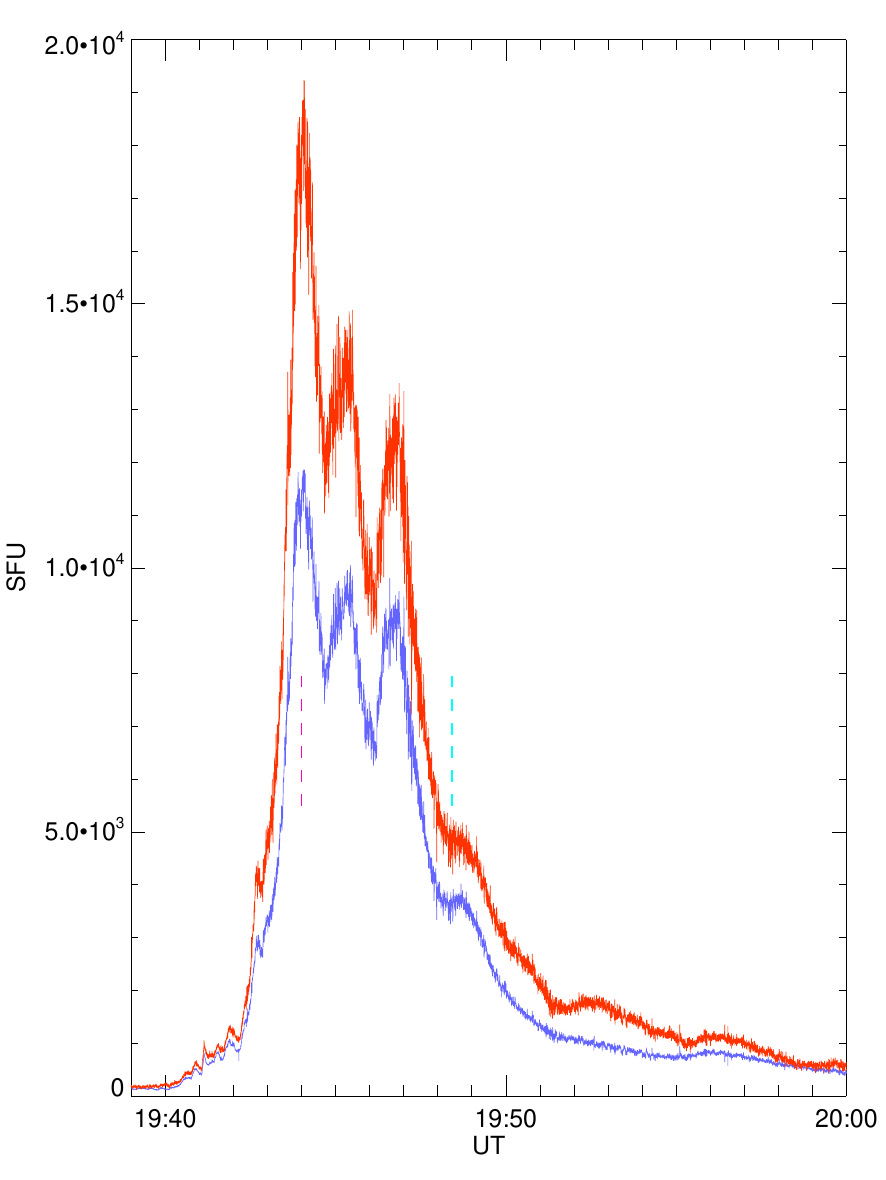}
  \includegraphics[width=40mm]{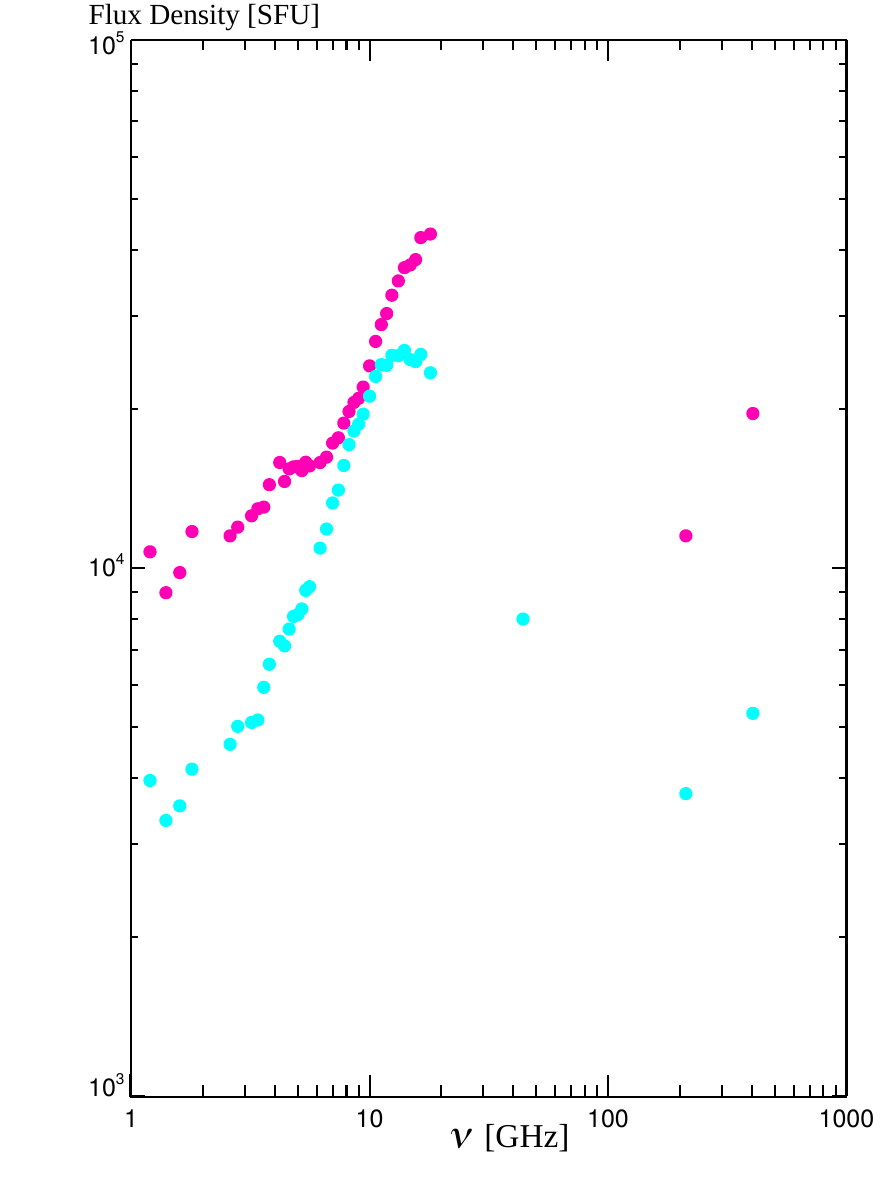}}
  \caption{Left: Time profiles of the SOL2003-11-04T1945 solar event
    at 212~GHz (blue) and 405~GHz (red) observed by the SST. Spectra
    obtained at two different instants of the event (see vertical
    dashed lines on the left panel). Microwave data were obtained by
    the OVSA array. Observations at 44~GHz were carried out at Pierre
    Kaufmann Radio Observatory (ROPK) with its 14-m single dish
    antenna. This was the first event to show a second spectral submm
    component \cite{Kaufmannetal:2004}.}
  \label{fig:sol20031104}
\end{figure}

Submillimeter-wave (submm) observations, here considered for $0.3 \le
\lambda \le 3$~mm, allow us to study the solar low atmospheric layers,
from the Transition Region to the Chromosphere
\cite{Wedemeyeretal:2016}. During flares, the submm emission might be
originated by synchrotron emission from relativistic particles
\cite{Ramatyetal:1994}. Therefore, we can track the energy transport
from the acceleration to the emission sites. Moreover, Kaufmann et
al. \cite{Kaufmannetal:2004} have shown that some flares have a second
spectral submm component (Figure \ref{fig:sol20031104}) with a still
unknown origin \cite{Kruckeretal:2013}.\\

Even though the immense wealth of information that submm
observations may bring to the understanding of the solar atmosphere
and its dynamics, there is a lack of regular observations to cover
this wavelength range. First efforts were carried on with the
James Clerk Maxwell Telescope (JCMT) \cite{LindseyKopp:1995}. The
Swiss KOSMA telescope, also observed the Sun a few times in 2003/2004
before it was decommissioned
\cite{Luthietal:2004a,Luthietal:2004b}. More recently, the Atacama
Large Millimeter Array (ALMA) is revealing fine details of the quiet
and quiescent solar behavior \cite{Wedemeyeretal:2020}. However, JCMT
and KOSMA observed the Sun just a couple of times, and ALMA allocates
a small portion of its observing time to the Sun and it is not the
best instrument to catch fast transient phenomena, like solar
flares. \\

Since 1999, the only solar dedicated submm instrument is the Solar
Submillimeter Telescope (SST) \cite{Kaufmannetal:2008}, a single dish
telescope with room temperature receivers operating at 212~GHz
($\lambda = 1.4$~ mm) and 405~GHz ($\lambda = 0.7$~mm). After more
than 20 years of excellent service, SST has to be updated in order to
provide answers to the questions its observations have raised: what is
the emission mechanism that creates a second spectral component $>
100$~GHz during flares? Does this component existe in ``weak'' flares? It
may also bring more information about the 3--5 minute p-mode
oscillations, the time evolution of the large scale chromospheric
structures and its relationship with the magnetic field, the ``slow''
components at these frequencies, among other. \\

In the following lines we will present the general characteristics of
the SST and introduce the SST next generation (\sstng) which is being
designed at the Center for Radio Astronomy and Astrophysics Mackenzie
(CRRAM) in São Paulo (Brazil).

\section{SST}

\begin{figure}[htbp]
  \centering
  \includegraphics[width=69mm]{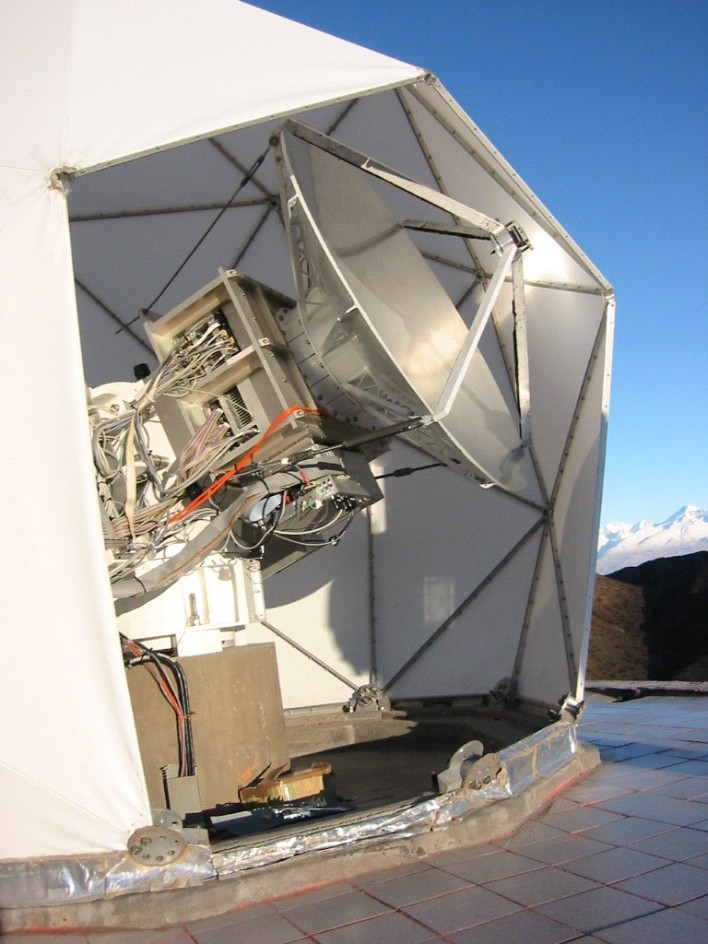}
  \caption{The SST with the radome open for maintenance.}
  \label{fig:sst}
\end{figure}

SST (Figure \ref{fig:sst}) is a product of \textit{state-of-art}
technologies of the 1990s. It has a 1.5~m, $f/D = 8$, radome-enclosed
single-dish aluminum reflector built at the Steward Observatory,
University of Arizona, Tucson, USA. Its frontend has six room
temperature radiometers that operate simultaneously: four receivers
operate at 212~GHz and two at 405~GHz, with nominal beam sizes of 4
and 2 arc minutes, respectively. The six beams form two arrays
separated by approximately 6 arc-minutes. The first array has three
212~GHz beams arranged in an equilateral triangle, in the center of
this triangle, there is one 405~GHz beam. The receiver horns have a
tapper that allows the beam intersection at 50\% level (-3
dB). Although the tapper reduces the efficiency and increases the
spill over, it allows the use of the \textit{multibeam} method to
instantly localize the emission centroid of point-like sources and to
correct the flux for offset pointing
\cite{GimenezDeCastroetal:1999,Costaetal:1995}. The second array has
one 212~GHz and one 405~GHz beam with the same center and is used for
reference. The radiometers have a $\Delta\nu=8.5$~GHz bandwidth,
temperatures of around 2000--3000~K and were custom made by
RPG-Radiometer GmbH, Meckenheim, Germany. The backend output signal is
converted to 2-byte integer numbers. SST has an Alt-Azimuth mount with
3.6~milliarcsec resolution and maximum speed of
$3^\circ\ \mathrm{s}^{-1}$. The output data is recorded in three
different file structures: \textit{sub-integrated} with 5 ms time
resolution, \textit{integrated} with 40~ms time resolution and
\textit{auxiliary} with 1~s time resolution.\\

The telescope is installed in the El Leoncito Astronomical Complex
(CASLEO, in Spanish) at 2550 m above sea level in the Argentinean
Andes, Province of San Juan. First light was in July 1999 and since April
2001 does daily observations. During the past 20+ years we refined the
measurement of the atmospheric optical depth with different techniques
and gathered a large statistics to understand the atmospheric
transmission on the site at both frequencies
\cite{Cassianoetal:2018,ValleSilvaetal:2019b}. Median values of the
opacity are 0.16 and 1.1 for 212 and 405 GHz, respectively
\cite{CornejoEspinozaetal:2020}. That means that for more than 50\% of
the time, the atmosphere is nominally optically thick at 405~GHz.\\

\begin{figure}[htbp]
  \centering
  \centerline{\includegraphics[width=65mm]{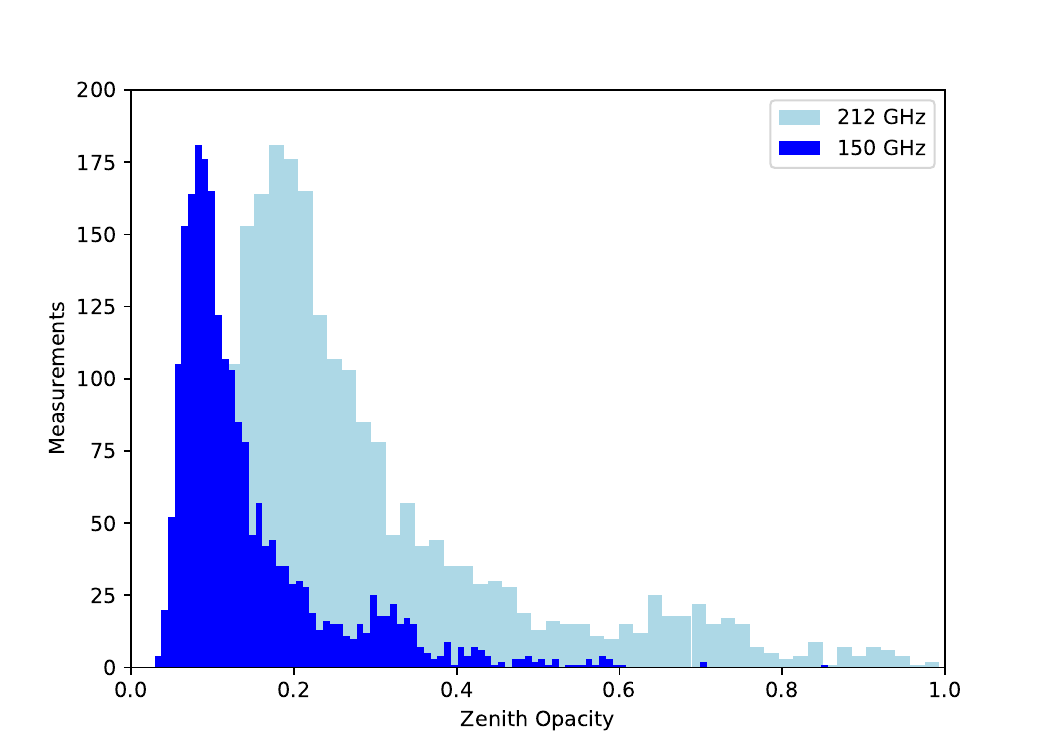}}
  \centerline{\includegraphics[width=65mm]{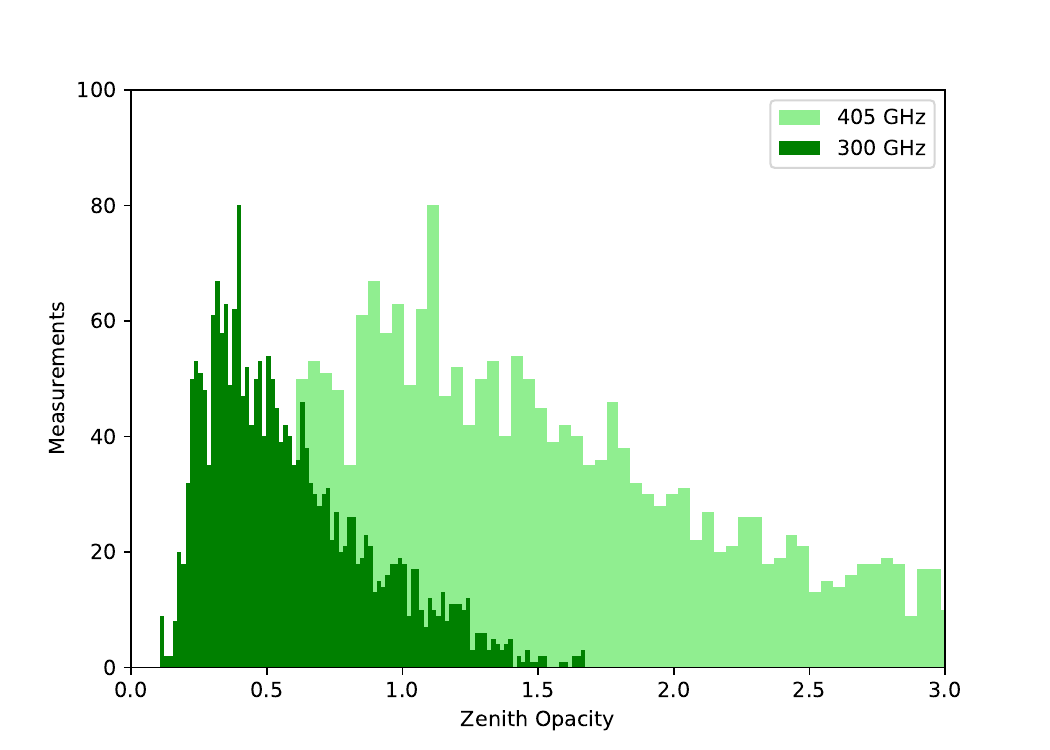}}
  \caption{Top panel: Observed atmospheric opacity histogram at
    212~GHz and expected opacity at 150~GHz.  Bottom panel: Observed
    opacity histogram at 405~GHz and expected opacity for 300~GHz. The
    observed opacities were determined with skydips between 2008 and
    2012.}
  \label{fig:tau}
\end{figure}

\section{\sstng}

The \sstng\ will be more sensitive. Indeed, the SST noise flux
density, when observing the Sun, is 1 and 7 SFU\footnote{Solar Flux
  Unit $\equiv 10^4$~Jy.} for 212 and 405 GHz, respectively,
considering 40~ms integration time, the median values of the
atmospheric opacities and a mean elevation angle of 60$^\circ$. With
this sensitivity, the weakest flares we have observed are of GOES class
M. By changing the receiver frequencies to 150 and 300~GHz we gain a
factor $> 2$ in opacity: from our statistics and using the
relationship obtained in \cite{ValleSilvaetal:2019b} we derive median
values of 0.07 and 0.4 for the atmospheric opacities at 150 and
300~GHz, respectively. In Figure \ref{fig:tau} we show the histograms
of the observed zenith opacities obtained between 2008 and 2012 using
the skydip method. The same figure presents the expected histograms for
150 and 300~GHz showing an expressive reduction.\\

We also want to keep the same beam sizes, therefore we plan to
substitute the present reflector by a new one of 3-m
diameter. Moreover, today receivers have lower temperatures and larger
bandwidths. For the present work we assume temperatures around 1000
and 2500~K for 150, and 300~GHz, respectively, and $\Delta\nu=16$~GHz
for both frequency bands. Everything combined, lower opacities, larger
reflector surface and receiver bandwidth, and smaller temperature,
shall yield noise fluxes of 0.06 and 0.12~SFU, i.e. \sstng\ will be 15
and 55 times more sensitive when compared with SST 212 and 405~GHz
observations. In terms of flares, this gain
means that events of GOES class C, and maybe, class B, will be
detected, dramatically increasing the number of events to analyze. In
terms of quiet sun behavior, it will certainly be possible to detect
the 3-5 minute oscillations, and faint structures. \\

Polarization is key to discriminate the origin of the emission and to
study the ambient magnetic field, however it was not yet explored at
submm wavelengths during flares. \sstng\ will be the first solar
telescope to have circular polarization detectors for both frequency
bands. And we plan to have a spectrometer to make studies of the yet
to be observed large $n$ Rydberg hydrogen lines at these
frequencies. On the other hand the multibeam system will be maintained
with three receivers at 300~GHz and one at 150~GHz in a triangular
array similar to SST.\\

\sstng\ will be able to make observations of non-solar objects like
H~\textsc{ii} regions and QSOs. Indeed, for 1-min integration time,
the noise flux density of \sstng\ will be 3 and 12~Jy for 150 and
300~GHz, respectively, making  possible night surveys.\\

\begin{figure}[htbp]
  \centering
  \centerline{\includegraphics[width=180mm, angle=-90]{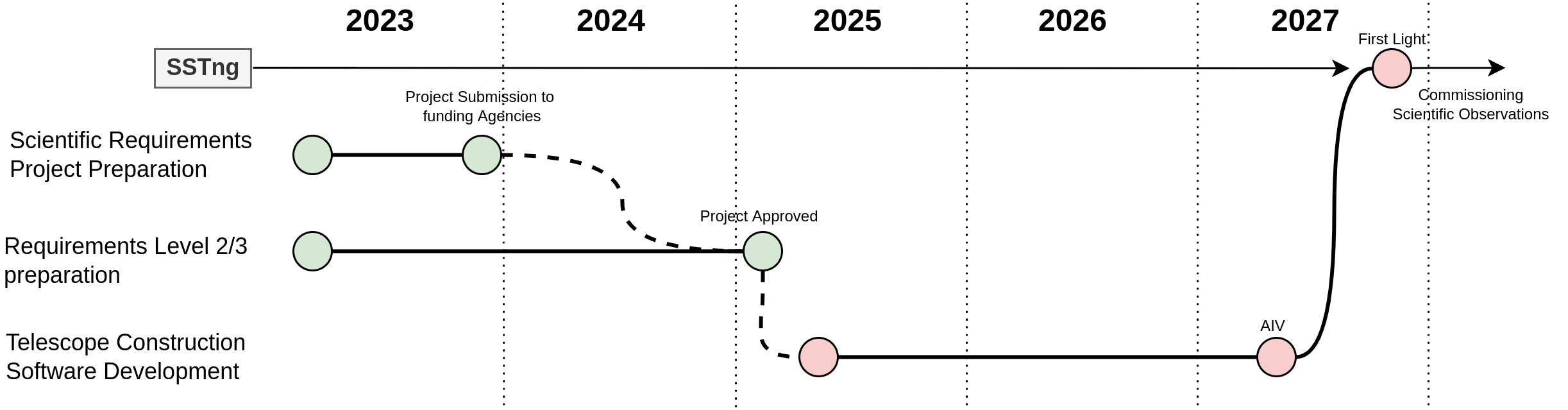}}
  \caption{Provisional project schedule.}
  \label{fig:cronograma}
\end{figure}

\section{Final remarks}

As we said above, \sstng\ is more than an update of the 1990s
technology. It intends to be a new instrument, based on our experience
in this frequency range that will enlarge the original scientific
goals. At the present time we are finishing the scientific
requirements, afterwards we will start to identify possible
contractors for the different subsystems. By the end of 2023 we will
submit projects to our funding agencies to obtain financial
support. Construction should start early 2025 and by 2027 it should
have its first light, starting the commissioning and the scientific
observations (Figure \ref{fig:cronograma}).

\section*{Acknowledgements}

We acknowledge FAPESP and CAPES funding agencies through grants
2013/24155-3 and 88887.310385/2018-00, respectively, for their support
to this scientific project.

\bibliographystyle{IEEEtran}

\end{document}